\documentclass[superscriptaddress,twocolumn,prl,showpacs]{revtex4}

\usepackage{dcolumn}
\usepackage{amsfonts}
\usepackage{amsmath}
\usepackage{amssymb}
\usepackage{bm}
\usepackage{epsfig}

\begin{document}

\newcommand{\brm}[1]{\bm{{\rm #1}}}
\newcommand{\tens}[1]{\bm{#1}}
\newcommand{\mm}{\overset{\leftrightarrow}{m}}
\newcommand{\xv}{\bm{{\rm x}}}
\newcommand{\Rv}{\bm{{\rm R}}}
\newcommand{\uv}{\bm{{\rm u}}}
\newcommand{\nv}{\bm{{\rm n}}}
\newcommand{\Nv}{\bm{{\rm N}}}
\newcommand{\ev}{\bm{{\rm e}}}
\newcommand{\phit}{\tens{\phi}}
\newcommand{\Tr}{\text{Tr}}

\title{Semi-soft Nematic Elastomers and Nematics in Crossed Electric and Magnetic Fields}

\author{Fangfu Ye}
\affiliation{Department of Physics and Astronomy, University of
Pennsylvania, Philadelphia, PA 19104, USA }

\author{Ranjan Mukhopadhyay}
\affiliation{Department of Physics, Clark University, Worcester, MA
01610, USA}

\author{Olaf Stenull}
\affiliation{Fachbereich Physik, Universit\"{a}t Duisburg-Essen,
Lotharstr.~1, 47048 Duisburg, Germany}

\author{T. C. Lubensky}
\affiliation{Department of Physics and Astronomy, University of
Pennsylvania, Philadelphia, PA 19104, USA }

\vspace{10mm}
\date{\today}
\begin{abstract}
Nematic elastomers with a locked-in anisotropy direction exhibit
semi-soft elastic response characterized by a plateau in the
stress-strain curve in which stress does not change with strain.
We calculate the global phase diagram for a minimal model, which
is equivalent to one describing a nematic in crossed electric and
magnetic fields, and show that semi-soft behavior is associated
with a broken symmetry biaxial phase and that it persists well
into the supercritical regime. We also consider generalizations
beyond the minimal model and find similar results.
\end{abstract}
 \maketitle

\begin{figure}
\centerline{\includegraphics[width=7cm]{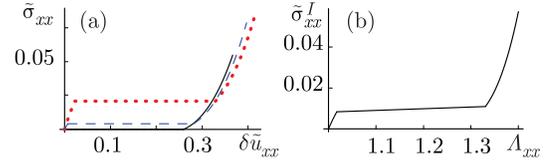}} \caption{(a):
Soft (full line) and semi-soft (dashed and dotted lines)
stress-strain curves at $\tilde{r}=0.08$ with $\tilde{h}=0,~ 0.8
\tilde{h}_c, ~4 \tilde{h}_c$, respectively.  (b) Semi-soft curve of
$\tilde{\sigma_{xx}}^I$ as a function of $\Lambda_{xx}$ at
$\tilde{r}=0.08$ and $\tilde{h}=2 \tilde{h}_c$, where we have set
$v=w$.}
\label{fig:simple_stress_strain}
\end{figure}

Nematic elastomers (NEs) \cite{WarnerTer2003} are remarkable
materials that combine the elastic properties of rubber with the
orientational properties of nematic liquid crystals.  An ideal
uniaxial nematic elastomer is produced when an isotropic rubber,
formed by crosslinking a polymer with nematogenic mesogens,
undergoes a transition to the nematic phase in which it
spontaneously stretches along one direction (the $z$-direction) and
contracts along the other two while its nematic mesogens align on
average along the stretch direction. This ideal nematic phase
exhibits soft-elasticity \cite{WarnerTer1994,Olmsted1994} -- a
consequence of Goldstone modes arising from the breaking of the
continuous rotational symmetry of the isotropic phase
\cite{GolubovicLub1989a}. Soft elasticity is characterized by the
vanishing of the elastic modulus $C_5$ measuring the energy
associated with shears $u_{xz}$ and $u_{yz}$ in planes containing
the anisotropy axis and by a stress-strain curve for strains
$u_{xx}$ (or $u_{yy}$) and stresses $\sigma_{xx}$ (or $\sigma_{yy}$)
perpendicular to the anisotropy axis in which strains up to a
critical value are produced at zero stress as shown in
Fig.~\ref{fig:simple_stress_strain}(a).

Monodomain samples cannot be produced without locking in a preferred
anisotropy direction, usually by the K\"{u}pfer-Finkelmann (KF)
procedure \cite{KupferFin1991} in which a first crosslinking in the
absence of uniaxial stress is followed by  second one with stress.
This process introduces a mechanical aligning field $h$, analogous
to an external electric or magnetic field, and lifts the value of
the elastic modulus $C_5$ from zero. Thus, nematic elastomers
prepared in this way are simply uniaxial solids with a linear stress
strain relation at small strain. For fields $h$ that are not too
large, however, they are predicted to exhibit semi-soft elasticity
\cite{WarnerTer2003,VerwayWar1995} in which the nonlinear
stress-strain curve exhibits a flat plateau at finite stress as
shown in Fig.~\ref{fig:simple_stress_strain}(a). Measured
stress-strain curves in appropriately prepared samples unambiguously
exhibit the characteristic semi-soft plateau
\cite{KupferFin1994,Warner1999}.

The Goldstone argument for soft response predicts $C_5 = 0$ in the
nematic phase, making reasonable conjectures that $C_5$ should
remain small at finite $h$ when semi-soft response is expected and
that semi-soft response might not exist at all in the supercritical
regime \cite{StenullLub2004-2} beyond the mechanical critical point
(with $h=h_c$) terminating the paranematic($PN$)-nematic($N$)
coexistence line \cite{deGennes1975-1}. There is now strong evidence
\cite{LebZal2005,RogezMar2006} that samples prepared with the KF
technique are supercritical. In addition, $C_5$ measured in
linearized rheological experiments is not particularly small
\cite{RogezMar2006}. These results have caused some to doubt the
interpretation of the measured stress-strain plateau in terms of
semi-soft response \cite{BrandMar2006}.

\begin{figure}
\centerline{\includegraphics[width=7.5cm]{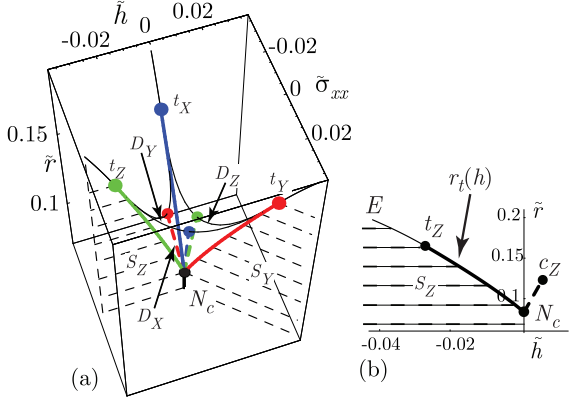}}
\caption{Phase diagrams (a) in the
$\tilde{h}$-$\tilde{\sigma}_{xx}$-$\tilde{r}$ space showing the
$S_Y$ and $S_Z$ ($S_X$ hidden)$CC$ and the $D_X$, $D_Y$, and $D_Z$
$DC$ surfaces along with the tricritical points $t_X$, $t_Y$, $t_Z$
and (b)in the $\tilde{r}$-$\tilde{h}$ plane
($\tilde{\sigma}_{xx}=0$) showing the first-order uniaxial $PN$-$N$
coexistence line $N_cc_Z$, the mechanical critical point $c_Z$, and
the $S_Z$ surface terminated by the line $\tilde{r}_t(\tilde{h})$
with respective first- and second-order segments $N_ct_Z$ and $t_ZE$
meeting at the tricritical point $t_Z$.} \label{fig:phase_dia1}
\end{figure}

The purpose of this paper is to clarify the nature of semi-soft
response.  We consider the simplest or minimal model, which is
formally equivalent to the Maier-Saupe-de-Gennes model
\cite{deGennesPro1994} for nematic liquid crystals, that exhibits
this response. We derive the global mean-field phase diagram
[Fig.~\ref{fig:phase_dia1}] for this model. We show that semi-soft
response is associated with biaxial phases that spontaneously break
rotational symmetry, and we unambiguously establish that semi-soft
response exists well into the supercritical regime. Figure
\ref{fig:simple_stress_strain} shows calculated stress-strain curves
for $h= 0.8 h_c$ and $h= 4 h_c$ that clearly exhibit semi-soft
behavior both for $h<h_c$ and in the supercritical regime with $h>
h_c$. Our minimal model provides a robust description of semi-soft
response. We will, however, briefly discuss changes in this response
that extensions of the minimal model can bring about.

An elastomer is characterized by an equilibrium reference
configuration, which we refer to as a reference space $S_R$, with
mass points at positions $\xv$. Upon distortion of the elastomer,
points $\xv$ are mapped to points $\Rv(\xv) \equiv \xv + \uv(\xv)$
in a target space $S_T$, where $\uv(\xv)$ is the displacement variable.
Elastic distortions that vary slowly on scales set by the distance
between crosslinks are described by the Cauchy deformation tensor
$\tens{\Lambda}$ with components $\Lambda_{ij} =\partial R_i
/\partial x_j$.  The usual Lagrangian strain tensor is then
$\tens{u} = (\tens{\Lambda}^T \cdot \tens{\Lambda} - \tens{\delta}
)/2$, where $\tens{\delta}$ is the unit matrix.  The orientational properties of nematic mesogens in the
elastomer are measured by the Maier-Saupe nematic tensor $\tens{Q}$.

A complete theory for nematic elastomers should treat both
$\tens{\Lambda}$ and $\tens{Q}$ and couplings between them. However,
effective theories, obtained by integrating out $Q_{ij}$, that
depend only on $\tens{u}$ provide a full description of the
mechanical properties of NEs
\cite{GolubovicLub1989a,LubenskyXin2002}. In such theories, strains
measure distortions relative to an isotropic reference state, and
the elastic free-energy density $f(\tens{u})$ consists of an
isotropic part $f_{\text{iso}} (\tens{u})$ and an anisotropic part
$f_{\text{ani}} (\tens{u}, h)$ arising from the imprinting process
\cite{KupferFin1991}. Equilibrium in the presence of an external
second Piola-Kirchhoff (PK) stress $\sigma_{xx}$ is determined by
minimization over $\tens{u}$ of the Gibbs free energy density
$g(\tens{u}, h, \sigma_{xx},r) = f_{\text{iso}} ( \tens{u},r )
+f_{\text{ani}}(\tens{u},h) + f_{\text{ext}}(\tens{u},\sigma_{xx})$,
where $f_{\text{ext}}(\tens{u},\sigma_{xx}, )=- \sigma_{xx} u_{xx}$.
In equilibrium the second PK stress satisfies $\sigma_{ij} =\partial
f/\partial u_{ij}$.  We will return later to the engineering or
first PK stress tensor $\sigma_{ij}^I = \partial f/\partial
\Lambda_{ij} = \Lambda_{ik} \sigma_{kj}$.

We now define our minimal model. First, we impose the constraint
$\Tr \tens{u} = 0$, enforcing incompressibility at small but not
large $\tens{u}$, rather than the full nonlinear incompressibility
constraint $\det \tens{\Lambda} = [\det (\tens{\delta} + 2 \tens{u})
]^{1/2} =1$ that more correctly describes NEs, whose bulk moduli are
generally orders of magnitude larger than their shear moduli.  Our
theory thus depends only on the symmetric-traceless components of
$\tens{u}$: $\phi_{ij} = u_{ij} - \frac{1}{3} \delta_{ij} u_{kk}$,
and $f_{\text{ext}} = - \sigma_{xx} \phi_{xx}$.  Second , we use the
simplest anisotropy energy: $f_{\text{ani}}  = - h u_{zz}
\rightarrow -h \phi_{zz}$ that favors stretching along the $z$ axis.
Thus, our theory is formally equivalent to that for a nematic liquid
crystal in crossed electric and magnetic fields, ${\bm {{\rm E}}} =
E \ev_z$ and ${\bm {{\rm H}}} = H \ev_x$,  in which $\phi_{ij}
\leftrightarrow Q_{ij}$, $h \leftrightarrow \frac{1}{2}{\Delta
\epsilon} E^2$, and $\sigma_{xx} \leftrightarrow \frac{1}{2} \chi_a
H^2$, where $\Delta \epsilon$ and $\chi_a$ are, respectively, the
anisotropic parts of the dielectric tensor and the magnetic
susceptibility, and $\ev_a$, $a=x,y,z$, are unit vectors along
direction $a$. Finally, we choose the Landau-de-Gennes form
\cite{deGennesPro1994} for $f_{\text{iso}}$:
\begin{equation}
f_{\text{iso}}(\phit,r) = \frac{1}{2} r \Tr \phit^2 - w \Tr \phit^3
+ v (\Tr \phit^2 )^2 , \label{eq:MSdG}
\end{equation}
where we assume $w>0$ and where $r = a(T-T^*)$ with $T$ the
temperature and $T^*$ the temperature at the metastability limit of
the $PN$ phase. In the isotropic phase with $\phit = 0$, $r= 2\mu$,
where $\mu$ is the $T$-dependent shear modulus. We will often
express quantities in reduced form: $\tilde{u}_{ij} = (v/w) u_{ij}$,
${\tilde r} = r v/w^2$, ${\tilde h} = h v^2/w^3$,
$\tilde{\sigma}_{ij} = \sigma_{ij} v^2/w^3$, $\tilde{C}_5 = C_5
v/w^2$, and similarly for other elastic moduli.

We begin our analysis of the global phase diagram~\cite{FriBerPal1987} with the
$\sigma_{xx} = 0$ plane, which we will refer to as the $Z$-plane
because the anisotropy field $h$ favors uniaxial order along the
$z$-axis. The $h\ge0$ half of this plane exhibits the familiar
nematic clearing point $N_c$ at $(\tilde{r}_N, \tilde{h}_N)=
(\frac{1}{12}, 0)$ and the $PN$-$N$ coexistence line terminating at
the mechanical critical point $(\tilde{r}_c, \tilde{h}_c)
=(\frac{1}{8}, \frac{1}{192})$.  Throughout the $h>0$ half-plane,
there is prolate uniaxial order with $\phi_{ij} = S(n_i n_j -
\frac{1}{3} \delta_{ij})$ with $S>0$ and the Frank director $\nv$
along $\ev_z$. In the $N$ phase at $h=0$ and $r<r_N$, $\nv$ can
point anywhere on the unit sphere. Negative $h$ induces oblate
rather than prolate uniaxial order along $\ev_z$ and $S = - S' <0$
at high temperature. When $h<0$ is turned on for $r<r_N$ at which
nematic order exists at $h=0$, $\nv$ aligns in the two-dimensional
$xy$-plane. This creates a biaxial environment and biaxial rather
than uniaxial order.  Since $\nv$ can point anywhere in the
$xy$-plane, the biaxial state at $h<0$ exhibits a spontaneously
broken symmetry.  There must be a transition along a line $r =
r_t(h)$ between the high-temperature oblate uniaxial state and the
low-temperature biaxial state, which exists throughout the $S_Z$
surface shown in Fig.~\ref{fig:phase_dia1}. This transition is first
order at small $|h|$ because the $PN$-$N$ transition is first order
at $h=0$ and second order at larger $|h|$, and there is a
tricritical point \cite{VargaSza2000} $t_Z$ at $(\tilde{r}_t,
\tilde{h}_t) = (\frac{21}{128},-\frac{27}{1024})$ separating the two
behaviors as shown in Fig.~\ref{fig:phase_dia1}(b). A continuum of
biaxial states coexist on $S_Z$.  We will refer to such surfaces as
$CC$ surfaces and ones on which a discrete set of states coexist as
$DC$ surfaces.

The full phase diagram reflects the symmetries of $g$. The $x$-
and $z$-directions are equivalent in $f_{\text{iso}}$, and the
$\sigma_{xx} = 0$ and the $h=0$ planes are symmetry equivalent.
These planes are also equivalent (apart from stretching) to the
vertical plane with $\sigma_{xx} = h$, but with positive and
negative directions interchanged.  To see this, we note that
$\phi_{zz} + \phi_{xx} = - \phi_{yy}$ and $h \phi_{zz} +
\sigma_{xx} \phi_{xx} = - h \phi_{yy}$ when $h= \sigma_{xx}$. Thus
the phase structure of the $Z$-plane is replicated in the
$X$-plane ($h = 0$) and the $Y$-plane ($\sigma_{xx} = h$) with
respective preferred uniaxial order along $\ev_x$ and $\ev_y$,
critical points $c_X$ and $c_Y$, biaxial coexistence surfaces
$S_X$ and $S_Y$, and tricritical points $t_X$ and $t_Y$.

To fill in the $3D$ phase diagram, we consider perturbations away
from the $X$-, $Y$-, and $Z$-planes.  Turning on $\sigma_{xx}$
converts the $PN$-$N$ coexistence line into a $DC$ surface $D_Z$,
on which two discrete in general biaxial phases coexist.  Turning
on $\sigma_{xx}$ near the $S_Z$ surface favors alignment of the
biaxial order along $\ev_x$ when $\sigma_{xx} >0$ and along
$\ev_y$ when $\sigma_{xx}<0$.  Thus $\sigma_{xx}$ is an ordering
field for biaxial order whereas a linear combination of $h$ and
$\sigma_{xx}$ acts as a nonordering field. The topology of the
phase diagram near $t_Z$ is that of the Blume-Emery-Griffiths
model \cite{BluGri1971} with $DC$ surfaces $D_X$ and $D_Y$
emerging from the first-order line $N_ct_Z$ terminating $S_Z$. The
$D_X$ and $D_Y$ surfaces terminate, respectively, on the critical
lines $N_ct_X$ and $N_ct_Y$ in the $X$- and $Y$-planes. The
surfaces $D_X$, $D_Y$, and $D_Z$ form a cone with vertex at $N_c$.

Before considering the $\sigma_{xx}$-$u_{xx}$ stress-strain curve,
it is useful to look more closely at elastic response in the
vicinity of the $Z$-plane and the nature of order in the $Y$-plane.
Throughout the $h>0$ $Z$-plane, the equilibrium state is prolate
uniaxial with order parameter $S=S_0$, and thus strains $u_{zz}^0  =
\frac{2}{3} S_0 =-2 u_{xx}^0 = -2 u_{yy}^0$. We are primarily
interested in shears in the $xz$-plane and the response to an
imposed $\sigma_{xx}$ with no additional stress along $z$. In this
case $\delta u_{zz} = u_{zz} - u_{zz}^0$ will relax to an imposed
$\delta u_{xx}$, and the free energy of harmonic deviations from
equilibrium can be written as $\delta f = \frac{1}{2} C_3 (\delta
u_{xx})^2 + \frac{1}{2} C_5 (\delta u_{xz})^2$.  The modulus $C_3$
gives the slope of $\sigma_{xx}$ versus $\delta u_{xx}$, and $C_5$
is measured in linearized rheology experiments
\cite{RogezMar2006,terentjev&Co_2003}. $C_3$ and $C_5$ are easily
calculated as a function of $r$ and $h$. In reduced units, the
ordered pair $(\tilde{C}_3, \tilde{C}_5)$ takes on the value
$(\frac{1}{8}, \frac{1}{6})$ just above $N_c$ ($\tilde{r} =
\tilde{r}_N^+$), $(\frac{3}{8},0)$ just below $N_c$ ($\tilde{r} =
\tilde{r}_N^-$), $(0,\frac{1}{12})$ at the critical point, and
$(\frac{57}{112}, \frac{1}{12})$ in the supercritical regime at
$(\tilde{r}, \tilde{h}) = (\tilde{r}_c, 2 \tilde{h}_c)$. We will
measure elastic distortions using $\delta u_{ij}$ rather than the
strain $u'_{ij}$ relative to the reference space $S'_R$ defined by
the equilibrium configuration at any given $T$ \cite{u'}.

On the $h>0$, $Y$-plane, there is oblate uniaxial order aligned
along the $y$-direction at high $T$ and biaxial order at low $T$.  A
convenient representation of the tensor order parameter is
\begin{equation}
\tens{\phi} = \left(
\begin{array}{ccc}
\frac{1}{3} S' - \eta_1 & 0 & \eta_2 \\
0 & -\frac{2}{3} S' & 0 \\
\eta_2 & 0 & \frac{1}{3} S' + \eta_1
\end{array}
\right ) , \label{eq:phit}
\end{equation}
where $S' >0$.  The vector $\vec{\eta} \equiv ( \eta_1, \eta_2)
\equiv \eta ( \cos 2\theta, \sin 2 \theta )$ is the biaxial order
parameter, which is nonzero on the $S_Y$ surface.  We define the
equilibrium values of $S'$ and $\eta$ in the biaxial phase to be
$S_0^{\prime}$ and $\eta_0$, respectively.  Energy in this phase is
independent of the rotation angle $\theta$.  Away from the
$Y$-plane, $f_{\text{ani}} + f_{\text{ext}}=  -\frac{1}{3} ( h +
\sigma_{xx} ) S' + (\sigma_{xx} - h)\eta_1$.  Thus, $\sigma_{xx} <h$
favors $\eta_1 >0$ and $\sigma_{xx} >h$ favors $\eta_1<0$, implying
that $\vec{\eta} = ( \eta_0,0)$ (or $\theta = 0$) at $\sigma_{xx} =
h^-$ and $\vec{\eta} = (- \eta_0 , 0 )$ (or $\theta = \frac{\pi}{2}$) at
$\sigma_{xx} = h^+$. These considerations imply that the modulus
$C_5$ is zero at $\sigma_{xx} = h^{\pm}$ because $C_5 =\partial^2 f
/\partial u_{xz}^2 |_{u_{xz}\rightarrow0}= (2\eta_0)^{-2}\partial^2
f/\partial \theta^2 |_{\theta\rightarrow0}= 0$.

We can now construct the $\sigma_{xx}$-$u_{xx}$ stress-strain
curves. At $\sigma_{xx} = 0$, $ u_{xx} = u_{xx}^0$; as $\sigma_{xx}$
is increased from zero, $\delta u_{xx}$ grows with initial slope
$1/C_3$ until $\sigma_{xx} = h^-$ at which point, $\delta u_{xx}
 = \frac{1}{3} S^{\prime }_0 - u_{xx}^0 - \eta_0$.
At $\sigma_{xx} = h$, further increase of $\delta u_{xx}$ to a
maximum of $\frac{1}{3} S^{\prime}_0 - u_{xx}^0 + \eta_0$ produces a
zero-energy rotation of $\vec{\eta}$ to yield $\delta u_{xx} =
\frac{1}{3} S^{\prime}_0 - u_{xx}^0 - \eta_0 \cos 2 \theta$ and a
nonzero shear $u_{xz} = \eta_2 = \eta_0 \sin 2 \theta$. The growth of
$\eta_2$ from zero is induced by the vanishing of $C_5$ at
$\sigma_{xx}= h^{\pm}$ and its becoming negative for $|\eta_1|<
\eta_0$.  Thus, the characteristic semi-soft plateau is a
consequence of $C_5$'s vanishing at $\sigma_{xx} = h$ and not at
$\sigma_{xx} = 0$. Measurements of $C_5$ at $\sigma_{xx} = 0$ do not
provide information about what happens at $\sigma_{xx} = h$. For
$\sigma_{xx}> h$, $\delta u_{xx}$ again grows with $\sigma_{xx}$.
Figure \ref{fig:simple_stress_strain} shows stress-strain curves for
different values of $\tilde h$. Thus, semi-soft response is
associated with the $S_Y$ surface, which exists at $r$ and $h$ well
into the supercritical regime.

A Ward identity provides a rigorous basis for the above picture
beyond mean-field theory. $f_{\text{iso}} (\tens{u})$ is invariant
under rotations of $\tens{u}$, i.e., under $\tens{u} \rightarrow
\tens{U} \tens{u}\tens{U}^{-1}$ where $\tens{U}$ is any rotation
matrix. Thus if $f_{\text{ani}}= -\Tr \tens{h} \tens{u}$, where
$h_{ij} = h e_{zi} e_{zj}$, $f(\tens{U} \tens{u} \tens{U}^{-1}) =
f_{\text{iso}} ( \tens{u}) - \Tr \tens h \tens{U}
\tens{u}\tens{U}^{-1}$,  for any $\tens{U}$, including one
describing an infinitesimal rotation by $\gamma$ about the $y$-axis
with components $U_{ij} = \delta_{ij} + \epsilon_{yij} \gamma$,
where $\epsilon_{ijk}$ is the Levi-Civita anti-symmetric tensor.
Equating the term linear in $\gamma$ in $f(\tens{U} \tens{u}
\tens{U}^{-1})$ to that of  $\Tr \tens h \tens{U} \tens{u}
\tens{U}^{-1}$ yields the Ward identity
\begin{equation}
\sigma_{xz} ( u_{zz} - u_{xx} ) = ( \sigma_{zz} + h - \sigma_{xx})
u_{xz} , \label{eq:ward}
\end{equation}
where $\sigma_{ij} = \partial f/\partial u_{ij}$.  This identity
applies for any $f_{\text{iso}}$, including ones with no
compressibility constraint, so long as $f_{\text{ani}}$ is linear in
$\tens{u}$. In the semi-soft geometry $\sigma_{xz} = \sigma_{zz} =
0$ but $\sigma_{xx} > 0$.  Thus, either $u_{xz} = 0$ or $\sigma_{xx}
= h$ for any nonzero $u_{xz}$. Equation (\ref{eq:ward}) also gives
$C_5 = \sigma_{xz}/u_{xz}|_{u_{xz} \rightarrow 0} = (h - \sigma_{xx}
)/(u_{zz} - u_{xx} ) = |h - \sigma_{xx} |/ 2 \eta_0$ implying that
$C_5 \rightarrow 0$ as $\sigma_{xx} \rightarrow h^{\pm}$ as long as
$\eta_0 \neq 0$.

\begin{figure}
\centerline{\includegraphics[width=7.5cm]{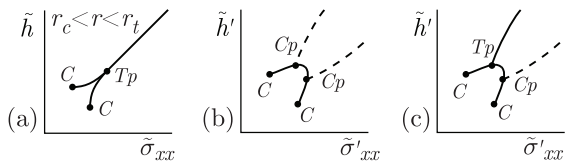}}
\caption{Schematic phase diagrams in the $h$(or $h'$)-stress plane.
The points $Tp$, $C$ and $Cp$ are, respectively, triple points,
liquid-gas-like critical points, and critical endpoints. (a)
Diagrams for the minimal model, where all transitions are first
order; (b) and (c) Phase diagrams for more general $f_{\text{ani}}$
or $f_{\text{ext}}$ in which the first-order line from $S_Y$ is
replaced by a surface terminated by two second-order (dashed) lines
or one first-order line and one second-order line. $h'$ and
$\sigma'_{xx}$ are, respectively, the generalized aligning field and
generalized external stress resulting from the more general
$f_{\text{ani}}$ or $f_{\text{ext}}$. } \label{fig:sigma_h_dia}
\end{figure}

We have focussed on the effects of an external second PK stress
$\sigma_{xx}$.  In physical experiments, the first PK (engineering)
stress, $\sigma_{ij}^I = \partial f/\partial \Lambda_{ij} =
\Lambda_{ik} \sigma_{kj}$, or the Cauchy stress, $\sigma_{ij}^C =
\sigma_{ik}^I \Lambda_{kj}^T/\det \tens{\Lambda}$ (as in
\cite{WarnerTer2003,Warner1999}), is externally controlled. The
$\sigma_{xx}^I$-$\Lambda_{xx}$ stress-strain curve is easily
obtained from the $\sigma_{xx}$-$u_{xx}$ curve using $\sigma_{xx}^I
= \Lambda_{xx} \sigma_{xx}$ and $\Lambda_{xx} = \sqrt{1 + 2
u_{xx}}$.  These two curves are similar, but the flat plateau in the
$\sigma_{xx}^I$-$\Lambda_{xx}$ curve rises linearly with
$\Lambda_{xx}$ as shown in Fig.~\ref{fig:simple_stress_strain}(b),
and there is a unique value of $\Lambda_{xx}$ for each value of
$\sigma_{xx}^I$. Thus, the $S_Y$ surface in the
$r$-$h$-$\sigma_{xx}$ phase diagram would open into a finite volume
biaxial region in the $r$-$h$-$\sigma_{xx}^I$ phase diagram with a
particular value of $\vec{\eta}$ at each point in it. The phase
diagram in the $h$-$\sigma_{xx}^I$ plane for $r_c<r<r_t$ is similar
to that in Fig.~\ref{fig:sigma_h_dia}(b).

We have ignored boundary conditions and random stress, both of which
can modify stress-strain curves. When Frank elastic energies are
ignored, detailed calculations of domain structure induced by
boundary conditions reproduce soft and semi-soft response
\cite{ContiDol2002a}.  Small isotropic randomness appears not to
affect soft response, but large randomness does \cite{Uchida2000}.
Our approach should serve as a basis for further study of
randomness.

We can now consider modifications of the minimal model.  A simple
modification is to replace the constraint $\Tr\tens{u}=0$ with the
real volume constraint $\det \tens{\Lambda} = 1$.  This replacement
does not change the validity of the Ward identity and the resulting
phase diagram has the same structure as that for $\Tr \tens{u} = 0$
but with different boundaries for the $CC$ and $DC$ surfaces.  In
particular, the mechanical critical point is at $(\tilde{r}_c,
\tilde{h}_c) = (0.1279,0.0052)$ and the tricritical point is at
$(\tilde{r}_t, \tilde{h}_t) = (0.1900,0.0247)$. Other modifications
of the minimal model replace $f_{\text{ani}}$ with nonlinear
functions of $u_{zz}$. Modifications of this kind can spread the
$CC$ surface $S_Y$ into a finite volume or convert it to a $DC$
surface, as shown in Fig. \ref{fig:sigma_h_dia}. If $f_{\text{ani}}=
-h u_{zz}^2$, two states coexist, whereas with other forms such as
might arise in a hexagonal lattice, three or more discrete states
might coexist. When $S_Y$ is a $DC$ surface, rather than exhibiting
a homogeneous rotation of the biaxial order parameter (if boundary
conditions are ignored) in response to an imposed $u_{xx}$, samples
will break up into disrete domains of the allowed states. In other
words, their response to external stress will be martensitic
\cite{Bhattacharya2003} rather than semi-soft.

The neo-classical model \cite{BlandonWar1994} can also be discussed
in our language.  The free energy of this model is a function of
$\tens{\Lambda}$ and $\tens{Q}$.  It consists of an isotropic part,
invariant under simultaneous rotations of $\tens{\Lambda}$ and
$\tens{Q}$ in the target space and under rotations of
$\tens{\Lambda}$ in the reference space, and a semi-soft anisotropic
energy \cite{VerwayWar1995}, which is effectively nonlinear in the
strain, that breaks rotational symmetry in the reference space. The
phase diagram of this model is similar to that of the minimal model in
the space of $r$-$h$-$\sigma_{xx}^I$.  In it, semi-soft behavior
also persists above the mechanical critical point \cite{YeLub2007}.

In summary, we determined the complete phase diagram of nematic
elastomers subject to an internal aligning field and a perpendicular
external stress. Our results underscore the validity of
semi-softness in the interpretation of their remarkable
stress-strain curves.

This work was supported by NSF grant DMR 0404570 and the NSF MRSEC
under DMR 05-20020.


\begin{thebibliography}{10}
\bibitem{u'} The deformation tensor of the anisotropic equilibrium
state at temperature $T$ relative to $S_R$ is  $\tens{\Lambda_0}$.
The strain and stress relative to $S_R'$ are, respectively,
$\tens{u}' = (\tens{\Lambda}_0^T)^{-1} \delta \tens{u}
(\tens{\Lambda}_0)^{-1}$ and $\tens{\sigma}' = (\det
\tens{\Lambda}_0)^{-1} \tens{\Lambda}_0 \tens{\sigma}
\tens{\Lambda}_0^T$.
\end{thebibliography}

\end{document}